\documentclass[runningheads]{llncs}

\usepackage[T1]{fontenc}
\usepackage[utf8]{inputenc}
\usepackage[english]{babel}
\usepackage{lmodern}
\usepackage{amsmath,tikz-cd}
\usepackage{amssymb}
\usepackage{oz}
\usepackage{hyperref}
\usepackage{graphicx}
\usepackage{xcolor}

\usepackage[linesnumbered,ruled]{algorithm2e}
\usepackage{listings}
\usepackage{subcaption}

\usepackage{ifthen}

\newcommand{\itv}[2]{[\![ #1, #2 ]\!]}
\newcommand{\card}[1]{|\!{#1}\!|}

\newcommand{\stateSet}{{\rm \Sigma}}
\newcommand{\oState}{{\tt \star}}
\newcommand{\gState}{{\tt G}}
\newcommand{\qState}{{\tt Q}}
\newcommand{\fState}{{\tt F}}
\newcommand{\aState}{{\tt A}}
\newcommand{\bState}{{\tt B}}
\newcommand{\cState}{{\tt C}}

\newcommand{\initConfigSet}{{\rm I}}
\newcommand{\initConfigFSSP}[1]{\overline{#1}}
\newcommand{\lConfig}[3]{(#1,#2,#3)}
\newcommand{\lTrans}[4]{\lConfig{#1}{#2}{#3} \mapsto #4}
\newcommand{\transTab}{{\rm \delta}}
\newcommand{\fDiagram}{{\rm D}}
\newcommand{\diagram}[4]{\fDiagram_{#1}({#2})({#3},{#4})}
\newcommand{\quintuplet}[5]{({#1},{#2},{#3},{#4},{#5})}
\newcommand{\triplet}[3]{\lConfig{#1}{#2}{#3}}
\newcommand{\superTransTab}{{\rm \Delta}}

\newcommand\family[1]{\mathrm{D}_{#1}}

\newcommand{\localSimFunc}[2]{f_{#1,#2}}
\newcommand{\candidate}[2]{\localSimFunc{#1}{#2}^{c}}

\newcommand{\neighborhood}{{N}}

\newcommand{\ie}{\emph{i.e.}~}

\newcommand{\mnote}[1]{\textcolor{red}{#1}}

\SetKwFunction{explore}{explore}
\SetKwFunction{perturbation}{perturbation}
\SetKwFunction{isSimul}{isSimul}
\SetKwFunction{pertN}{pertN}

\begin{document}

\title{Exploring Millions of 6-State FSSP Solutions: the Formal Notion of Local CA Simulation}
\titlerunning{Millions of 6-State FSSP Solutions through Local CA Simulations}
\author{
  Tien Thao Nguyen
  \and Luidnel Maignan
}
\authorrunning{T.T. Nguyen and L. Maignan}
\institute{
  LACL, Universit\'e Paris-Est Cr\'eteil, France
}
\maketitle

\begin{abstract}
In this paper, we come back on the notion of local simulation allowing to transform a cellular automaton into a closely related one with different local encoding of information.
This notion is used to explore solutions of the Firing Squad Synchronization Problem that are minimal both in time ($2n-2$ for $n$ cells) and, up to current knowledge, also in states (6 states).
While only one such solution was proposed by Mazoyer since 1987, 718 new solutions have been generated by Clergue, Verel and Formenti in 2018 with a cluster of machines.
We show here that, starting from existing solutions, it is possible to generate millions of such solutions using local simulations using a single common personal computer.
\end{abstract}


\section{Introduction}

\subsection{Firing Squad Synchronization Problem and the Less-State Race}

\label{intro}
The Firing Squad Synchronization Problem (FSSP) was proposed by John Myhill in 1957.
The goal is to find a single cellular automaton that synchronizes any one-dimensional horizontal array of an arbitrary number of cells.
More precisely, one consider that at initial time, all cells are inactive (i.e. in the \emph{quiescent state}) except for the leftmost cell which is in the general (i.e. in the \emph{general state}).
One wants the evolution of the cellular automaton to lead all cells to transition to a special state (i.e. the \emph{synchronization} or \emph{firing state}) \emph{for} the first time \emph{at} the same time.
This time $t_s$ is called the synchronization time and it is known that its minimal possible value is $2n - 2$ where $n$ is the number of cells.

For this problem, many minimum-time solutions were proposed using different approaches.
As indicated in~\cite{DBLP:journals/amc/UmeoHNIS18}, the first one was proposed by Goto in 1962~\cite{DBLP:journals/amc/UmeoHNIS18} with many thousands of states, followed by Waksman in 1966~\cite{DBLP:journals/iandc/Waksman66}, Balzer in 1967~\cite{DBLP:journals/iandc/Balzer67}, Gerken in 1987~\cite{10012108193}, and finally Mazoyer in 1987~\cite{DBLP:journals/tcs/Mazoyer87} who presented respectively a 16-state, 8-state, 7-state and 6-state minimum-time solution, with no further improvements since 1987.
Indeed, Balzer~\cite{DBLP:journals/iandc/Balzer67} already shows that there are no 4-state minimal-time solutions, latter confirmed by Sanders~\cite{DBLP:conf/parcella/Sanders94} through an exhaustive search and some corrections to Balzer's work.
Whether there exist any 5-state minimumal-time solution or not is still an open question.

Note that all these solutions use a ``divide and conquer'' strategy.
Goto's solution were pretty complex with two types of divisions.
The following ones used a ``mid-way'' division but Mazoyer's 6-state solution uses for the first time a ``two-third'' type of division\footnote{See Figure~\ref{fig:half} for a mid-way division, and Figure~\ref{fig:two-third} for a two-third division}.
In 2018, Clergue, Verel and Formenti~\cite{DBLP:journals/asc/ClergueVF18} generated 718 new 6-state solutions using an Iterated Local Search algorithm to explore the space of 6-state solutions on a cluster of heterogeneous machines:
717 of these solutions use a ``mid-way'' division, and only one use a ``two-third'' division.

\subsection{Context of the Initial Motivations}

In 2012, Maignan and Yunes proposed the methodology of \emph{cellular fields} to described formally the high-level implementation of a CA, and also formally the generation of the ``low-level'' transition table.
One of the expected benefits was to have an infinite CA cleanly modularized into many cellular fields with clear semantical proof of correctness together with a correctness-preserving reduction procedure into a finite state CA \cite{DBLP:conf/acri/MaignanY12} using a particular kind of cellular field, ``reductions''.
This is very similar to what happen in usual computer programming where one writes in a high-level language then transform the code into assembly using a semantic-preserving transformation, \ie a compiler.
In 2014, they made precise a particular reduction of the infinite CA into 21 states~\cite{DBLP:conf/acri/MaignanY14,DBLP:journals/jca/MaignanY16}.

From theses works and concepts, two intertwined research directions emerge.
One direction is to ask whether a reduction to fewer states is firstly possible, and secondly automatically generable, in the same spirit as compiler optimization, with the possible application of reducing further the 21 states.
The second direction is to build a map of as many FSSP solutions as possible and study how they relate through the notion of ``reduction'' introduced, with application the discovery of techniques used in hand-made transition table and also the factorisation of correctness proofs.
In 2018, Maignan and Nguyen~\cite{DBLP:journals/jca/NguyenM20} exhibited a few of these relations and in particular the fact the infinite Maignan-Yunes CA could be reduced to the 8-state solutions of Noguchi~\cite{DBLP:journals/tcs/Noguchi04}.

\subsection{From the Initial Motivations to a Surprise}

The initial motivations whose to complete the ``map of reductions'' by including the 718 solutions into the picture.
In particular, a quick look at the 718 solutions gave to the authors the feeling that they could be grouped into equivalence classes using the notion of ``reduction''.
Also, inspired by the idea of \emph{local search} and exploration through small modifications used in~\cite{DBLP:journals/asc/ClergueVF18} to generate the 718 solutions,
the first author tried such search algorithms to generate ``reductions'' of existing solutions rather than transition tables directly.
Although the idea of local search is to navigate randomly in a landscape with few actual solutions, the discovered landscape of reductions has so many solutions that a ``best-effort-exhaustive'' exploration have been tried, leading to \emph{many millions} of 6-states solutions.
Also, this space is much more easily explored because of its nice computational properties.

\subsection{Organization of the Content}

In Section~\ref{background}, we define formally cellular automata, local simulations, FSSP solutions and related objects.
In Section~\ref{sec:prop}, we present nice properties relating these objects and allowing the search algorithm to save a huge amount of time.
In Section~\ref{optim-problem}, we describe the exploration algorithm and continue in Section~\ref{result} with some experimental results and a small analysis of the 718 solutions.
We conclude in Section~\ref{conclusion} with some formal and experimental futur work.

\section{Background}
\label{background}

In this section, we define formally cellular automata, local mappings and FSSP solutions in a way suitable to the current study.
Some objects have ``incomplete'' counterpart manipulated during the exploration algorithm.
The material here is a considerable re-organization of the material found in~\cite{DBLP:journals/jca/NguyenM20}.

\subsection{Cellular Automata}

\begin{definition}\label{def:ca}\label{def:std}
    A \emph{cellular automaton} $\alpha$ consists of a finite set of \emph{states} $\stateSet_\alpha$, a set of \emph{initial configurations} $\initConfigSet_\alpha \subseteq {\stateSet_\alpha}^\mathbb{Z}$ and a partial function $\transTab_\alpha : {\stateSet_\alpha}^3 \pfun \stateSet_\alpha$ called the \emph{local transition function} or \emph{local transition table}.
    The elements of ${\stateSet_\alpha}^\mathbb{Z}$ are called \emph{(global) configurations} and those of ${\stateSet_\alpha}^3$ are called \emph{local configurations}.
    For any $c \in \initConfigSet_\alpha$, its \emph{space-time diagram} $\fDiagram_\alpha(c) : \mathbb{N} \times \mathbb{Z} \to \stateSet_\alpha$ is defined as:
    \begin{equation*}
        \diagram{\alpha}{c}{t}{p} =
        \begin{cases}
            c(p) & \text{ if } t = 0,\\
            \transTab_\alpha(l_{-1}, l_0, l_1) & \text{ if } t > 0 \text{ with } l_i = \diagram{\alpha}{c}{t-1}{p+i}.
        \end{cases}
    \end{equation*}
    The partial function $\transTab_\alpha$ is required to be such that all space-time diagrams have to be totally defined.
    When $\fDiagram_\alpha(c)(t,p) = s$, we say that, for the cellular automaton $\alpha$ and initial configuration $c$, the cell at position $p$ has state $s$ at time $t$.
\end{definition}

\begin{definition}\label{family-std}
    A \emph{family of space-time diagrams $D$} consists of a set of states $\stateSet_D$ and an arbitrary set $D \subseteq {\stateSet_D}^{\mathbb{N} \times \mathbb{Z}}$ of space-time diagram.
    The \emph{local transition relation $\delta_D \subseteq {\stateSet_D}^3 \times \stateSet_D$ of $D$} is defined as:
    $$((l^0_{-1}, l^0_0, l^0_1), l^1_0) \in \delta_D :\iff \exists(d,t,p) \in D \times \mathbb{N} \times \mathbb{Z} \text{ s.t. } l^j_i = d(t+j,p+i).$$
    We call $D$ a \emph{deterministic family} if its local transition relation is functional.
\end{definition}


\begin{definition}\label{def:family-to-ca}
    Given a deterministic family $D$, its \emph{associated cellular automaton} ${\rm \Gamma}_D$ is defined as having the set of states $\stateSet_{{\rm \Gamma}_D} = \stateSet_D$, the set of initial configurations $\initConfigSet_{{\rm \Gamma}_D} = \{ d(0,\--) \mid d \in D \}$\footnote{Here, $d(0,\--)$ is the function from $\mathbb{Z}$ to $S$ defined as $d(0,\--)(p) = d(0,p)$.}, and the local transition function $\transTab_{{\rm \Gamma}_D} = \delta_{D}$.
\end{definition}
\begin{definition}
    Given a cellular automaton $\alpha$, its \emph{associated family of space-time diagram} (abusively denoted) $\family{\alpha}$ is defined as having the set of states $\stateSet_{\family{\alpha}} = \stateSet_\alpha$, and the set of space-time diagram $\family{\alpha} := \{\,\fDiagram_\alpha(c) \mid c \in \initConfigSet_\alpha \,\}$ and is clearly deterministic.
\end{definition}

These inverse constructions shows that deterministic families and cellular automata are two presentations of the same object.
For practical purposes, it is also useful to note that, since $\transTab_D$ has a finite domain, there are finite subsets of $D$ that are enough to specify it completely.

\subsection{Local Mappings and Local Simulations}

Theses two concepts are more easily pictured with space-time diagrams.
Given a space-time diagram $d \in S^{\mathbb{N} \times \mathbb{Z}}$, we build a new one $d'$ by determining each state $d'(t,p)$ from a little cone $\langle d(t-dt,p+dp) \mid dt \in \{0,1\}, dp \in \itv{-dt}{+dt} \rangle$ in $d$.
This cone is simply a state for $t = 0$, and when $d$ is generated by a cellular automaton,  this cone is entirely determined by $\langle d(t-1,p+dp) \mid dp \in \itv{-1}{1} \rangle$ for $t \ge 1$.
Since the set of all these triplets is exactly $\dom(\transTab_\alpha)$, the following definitions suffice for the current study.
We call this a local mapping, because the new diagram is determined locally by the original one.
When transforming a deterministic family, the result might not be deterministic, but if it is, we speak of a local simulation between two CA.

\begin{definition}\label{local-mapping}
A \emph{local mapping $h$ from a CA $\alpha$ to a finite set $S$} consists of two functions $h_{\mathtt{z}}: \{ d(0,x) \mid (d, x) \in \family{\alpha} \times \mathbb{Z} \} \to S$ and $h_{\mathtt{s}}: \dom(\transTab_\alpha) \to S$.
\end{definition}

\begin{definition}\label{def:generated-family}
    Given a local mapping $h$ from a CA $\alpha$ to a finite set $S$, we define its \emph{associated family of diagrams} $\Phi_h = \{ h(d) \mid d \in \family{\alpha} \}$ where:
    \begin{equation*}
        h(d)(t,p) =
        \begin{cases}
            h_\mathtt{z}(d(0,p)) & \text{ if } t = 0,\\
            h_\mathtt{s}(l_{-1}, l_0, l_1) & \text{ if } t > 0 \text{ with } l_i = d(t-1)(p+i).
        \end{cases}
    \end{equation*}
\end{definition}

\begin{definition}\label{def:local-simulation}
    A local mapping $h$ from a CA $\alpha$ to a finite set $S$ whose associated family of diagrams $\Phi_h$ is deterministic is called a \emph{local simulation from $\alpha$ to ${\rm \Gamma}_{\Phi_h}$}.
\end{definition}

\begin{proposition}
    Equivalently, a \emph{local simulation $h$ from a CA $\alpha$ to a CA $\beta$} is a local mapping from $\alpha$ to the set $\stateSet_\beta$ such that $\{ h_{\mathtt{z}}(c) \mid c \in \initConfigSet_\alpha \} = \initConfigSet_\beta$ and for all $(c, t, p) \in \initConfigSet_\alpha \times \mathbb{N} \times \mathbb{Z}$, we have $h_{\mathtt{s}}(l_{-1}, l_0, l_1) = l'_0$ with $l_i = \diagram{\alpha}{c}{t}{p + i}$ and $l'_0 = \diagram{\beta}{h_{\mathtt{z}}(c)}{t + 1}{p}$.
    The details of these formula are more easily seen graphically.
\end{proposition}
\begin{center}
    \def\sc{.6}
    \def\dotsStyle{\large}
\def\mstep{9}
\begin{tikzpicture}[scale=\sc, every node/.style={minimum size=1.0cm}]

\node at (0.5,0.5) {\dotsStyle{$\vdots$}};
\node at (\mstep + 0.5,0.5) {\dotsStyle{$\vdots$}};
\foreach \y in {1.5, ..., 4.5} {
	\pgfmathtruncatemacro{\t}{3.5-\y};
	\node at (0.5,\y) {\scriptsize \ifthenelse{\t < 0}{$t\t$}{\ifthenelse{\t > 0}{$t+\t$}{$t$}}};
	\node at (\mstep + 0.5,\y) {\scriptsize \ifthenelse{\t < 0}{$t\t$}{\ifthenelse{\t > 0}{$t+\t$}{$t$}}};
}
\node at (0.5,5.5) {\dotsStyle{$\vdots$}};
\node at (\mstep + 0.5,5.5) {\dotsStyle{$\vdots$}};

\node[scale=\sc] at (1.5, 6.5) {\dotsStyle{$\dots$}};
\node[scale=\sc] at (\mstep + 1.5, 6.5) {\dotsStyle{$\dots$}};
\foreach \x in {2.5, ..., 6.5} {
	\pgfmathtruncatemacro{\p}{-4.5 + \x};
	\node[scale=\sc] at (\x, 6.5) {\ifthenelse{\p < 0}{$p\p$}{\ifthenelse{\p > 0}{$p+\p$}{$p$}}};
	\node[scale=\sc] at (\mstep + \x, 6.5) {\ifthenelse{\p < 0}{$p\p$}{\ifthenelse{\p > 0}{$p+\p$}{$p$}}};
}
\node[scale=\sc] at (7.5, 6.5) {\dotsStyle{$\dots$}};
\node[scale=\sc] at (\mstep + 7.5, 6.5) {\dotsStyle{$\dots$}};

\draw[step=1.0cm, color=black] (2.0,1.0) grid (7.0,5.0);
\draw[step=1.0cm, color=black] (\mstep + 2.0,1.0) grid (\mstep + 7.0,5.0);

\foreach \y in {1, ..., 5}{
	\draw[step=1.0cm, color=black, dashed] (1.5,\y) -- (2,\y);
	\draw[step=1.0cm, color=black, dashed] (\mstep + 1.5,\y) -- (\mstep + 2,\y);	
	\draw[step=1.0cm, color=black, dashed] (7,\y) -- (7.5,\y);
	\draw[step=1.0cm, color=black, dashed] (\mstep + 7,\y) -- (\mstep + 7.5,\y);	
}

\foreach \x in {2, ..., 7} {
	\draw[step=1.0cm, color=black, dashed] (\x,1) -- (\x,0.5);
	\draw[step=1.0cm, color=black, dashed] (\mstep + \x,1) -- (\mstep + \x,0.5);	
	\draw[step=1.0cm, color=black, dashed] (\x,5) -- (\x,5.5);
	\draw[step=1.0cm, color=black, dashed] (\mstep + \x,5) -- (\mstep + \x,5.5);	
}

\draw[very thick] (3,4) -- (6,4) -- (6,3) -- (5,3) -- (5,2) -- (4,2) -- (4,3) -- (3,3) -- (3,4);
\draw[very thick] (\mstep + 3,4) -- (\mstep + 6,4) -- (\mstep + 6,3) -- (\mstep + 5,3) -- (\mstep + 5,2) -- (\mstep + 4,2) -- (\mstep + 4,3) -- (\mstep + 3,3) -- (\mstep + 3,4);

\foreach \x in {-1, 0, 1} {
	\node at (4.5 + \x, 3.5) {$l_{\x}$};
}
\node at (\mstep + 4.5, 2.5) {$l'_0$};

\node at (4.5, -0.5) {$\fDiagram_\alpha(c)$};
\node at (\mstep + 4.5, -0.5) {$\fDiagram_\beta(h_{\mathtt{z}}(c))$};

\end{tikzpicture}

\end{center}

\subsection{The Firing Squad Synchronization Problem}
\label{fssp}


\begin{definition}\label{fssp_ready}
    A cellular automaton is \emph{FSSP-candidate} if there are four special states $\oState_\alpha, \gState_\alpha, \qState_\alpha, \fState_\alpha \in \stateSet_\alpha$, if $\initConfigSet_\alpha = \{ \initConfigFSSP{n}_\alpha \mid n \ge 2 \}$ with $\initConfigFSSP{n}_\alpha$ being the \emph{FSSP initial configuration} of size $n$, i.e. $\initConfigFSSP{n}(p) = \oState_\alpha$, $\gState_\alpha$, $\qState_\alpha$, $\oState_\alpha$ if $p$ is respectively $p \le 0$, $p = 1$, $p \in \itv{2}{n}$, and $p \ge n + 1$.
    Moreover, $\oState_\alpha$ must be the \emph{outside state}, \emph{i.e.} for any $(l_{-1},l_0,l_1) \in \dom(\transTab_\alpha)$, we must have $\transTab(l_{-1},l_0,l_1) = \oState_\alpha$ if and only if $l_0 = \oState_\alpha$.
    Also, $\qState_\alpha$ must be a \emph{quiescent state} so $\transTab_\alpha(\qState_\alpha,\qState_\alpha,\qState_\alpha) = \transTab_\alpha(\qState_\alpha,\qState_\alpha,\oState_\alpha) = \qState_\alpha$.
\end{definition}

The $\oState_\alpha$ state is not really counted as a state since it represents cells that should be considered as non-existing.
Therefore, a FSSP-candidate cellular automaton $\alpha$ will be said to have $s$ states when $|\stateSet_\alpha \setminus \{\oState_\alpha\}|\; = s$, and $m$ transitions when $|\dom(\transTab_\alpha) \setminus\stateSet_\alpha\times\{\oState_\alpha\}\times\stateSet_\alpha| = m$.

\begin{definition}\label{min_fssp_solution}
A FSSP-candidate cellular automaton $\alpha$ is a \emph{minimal-time FSSP solution} if for any size $n$, $\fDiagram_\alpha(\initConfigFSSP{n})(t,p) = \fState_\alpha$ if and only if $t \ge 2n - 2$ and $p \in \itv{1}{n}$.
We are only concerned with minimal-time solutions but sometimes simply write \emph{FSSP solution}, or \emph{solution} for short.
\end{definition}


\section{Some Useful Algorithmic Properties}
\label{sec:prop}

Our global strategy to find new FSSP solutions is to build them from local simulations of already existing FSSP solution $\alpha$.
Taking the previous definitions litteraly could lead to the following procedure for a given local mapping $h$.
First, generates as many space-time diagrams of $\family{\alpha}$.
Secondly, use $h$ to transform each diagram $d \in \family{\alpha}$ into a new one $h(d)$, thus producing a sub-family of $\Phi_h$.
At the same time, build $\transTab_{\Phi_h}$ by collecting all local transitions appearing in each $h(d)$ and check for determinism and correct synchronization.
If every thing goes fine, we have a new FSSP solution $\beta = {\rm \Gamma}_{\Phi_h}$.

Such a procedure is time-consuming.
We show here useful properties that reduces drastically this procedure to a few steps.
In fact, the space-time diagrams of $\Phi_h$ never needs to be computed, neither to build the local transition relation $\delta_{\Phi_h}$ (Section~\ref{sltt}), nor to check that ${\rm \Gamma}_{\Phi_h}$ is an FSSP solution as showed in this section (Section~\ref{sec:correctness-preserving}).

\subsection{Summarizing Families into Super Local Transition Tables}
\label{sltt}


When trying to construct a CA $\beta$ from a CA $\alpha$ and a local mapping $h$ from the families of space-time diagrams as suggested by the formal definitions, there is huge amount of redundancy.
All entries of the local transition relation $\transTab_{\Phi_h}$ appear many times in $\Phi_h$, each of them being produced from the same recurring patterns in the space-time diagrams of $\alpha$.
In fact, it is more efficient to simply collect these recurring patterns that we may call \emph{super local transitions}, and work from them without constructing $\Phi_h$ at all.
It is specially useful because we consider a huge number of local mappings from a single CA $\alpha$.

\begin{definition}
    For a given CA $\alpha$, the \emph{super local transition table} $\Delta_\alpha$ consists of two sets $(\Delta_\alpha)_\mathtt{z} \subseteq {\stateSet_\alpha}^{3}$ and $(\Delta_\alpha)_\mathtt{s} \subseteq {\stateSet_\alpha}^{5} \times {\stateSet_\alpha}^{3}$ defined as:
    \begin{align*}
        (s_{-1},s_0,s_1)\phantom{)} \in (\Delta_\alpha)_\mathtt{z} :\iff \exists 
        & (d, p) \in \family{\alpha} \times \mathbb{Z} \\
        & \text{ s.t. } s_i = d(0, p+i), \\
        ((s^0_{-2},s^0_{-1},s^0_0,s^0_1,s^0_2), (s^1_{-1},s^1_0,s^1_1)) \in (\Delta_\alpha)_\mathtt{s} :\iff \exists 
        & (d, t, p) \in \family{\alpha} \times \mathbb{N} \times\mathbb{Z} \\
        & \text{ s.t. } s^j_i = d(t+j, p+i)
    \end{align*}
\end{definition}

Once all these patterns collected, it is possible to construct the local transition relation $\transTab_{\Phi_h}$ as specified in the following proposition.

\begin{proposition}
Let $h$ be a local mapping from a CA $\alpha$ to a set $S$.
The local transition relation $\transTab_{\Phi_h}$ of the family of space-time diagram $\Phi_h$ generated by $h$ and the super local transition function $\Delta_\alpha$ of $\alpha$ obey:
\begin{eqnarray*}
((l^0_{-1},l^0_0,l^0_1), l^1_0) \in \transTab_{\Phi_h}
& \iff & \exists (s_{-1},s_0,s_1)\phantom{)} \in (\Delta_\alpha)_\mathtt{z} \\
& & \text{ s.t. } l^0_i = h_\mathtt{z}(s_{i}) \text{ and } l^1_0 = h_\mathtt{s}(s_{i-1},s_{i},s_{i+1}) \\
& \vee & \exists ((s^0_{-2},s^0_{-1},s^0_0,s^0_1,s^0_2), (s^1_{-1},s^1_0,s^1_1)) \in (\Delta_\alpha)_\mathtt{s}\\
& & \text{ s.t. } l^j_i = h_\mathtt{s}(s^j_{i-1},s^j_{i},s^j_{i+1})
\end{eqnarray*}
\end{proposition}

We know have an efficient way to build the local transition relation $\transTab_{\Phi_h}$.
When it is functional, it determines a cellular automaton $\beta = {\rm \Gamma}_{\Phi_h}$.
For our purpose, we need to test or ensure in some way that $\beta$ is an FSSP solution.

\subsection{Local Mappings and FSSP}
\label{sec:correctness-preserving}

We first note that the constraints put by the FSSP on space-time diagrams induces constraints on local simulations between FSSP solutions.
So we can restrict our attention to local mappings respecting these constraints as formalized by the following definition and proposition.

\begin{definition}\label{fssp_compliant}
   A local mapping $h$ from a FSSP solution $\alpha$ to the states $\stateSet_\beta$ of a FSSP-candidate CA $\beta$ is said to be \emph{FSSP-compliant} if it is such that
   (0) $h_\mathtt{z}$ maps $\oState_\alpha$, $\gState_\alpha$, and $\qState_\alpha$ respectively to $\oState_\beta$, $\gState_\beta$, and $\qState_\beta$,
   (1) $h_\mathtt{s}\lConfig{l_{-1}}{l_0}{l_1} = \oState_\beta$ if and only if $\delta_\alpha\lConfig{l_{-1}}{l_0}{l_1} = \oState_\alpha$ (meaning simply $l_0 = \oState_\alpha$),
   (2) $h_\mathtt{s}\lConfig{l_{-1}}{l_0}{l_1} = \fState_\beta$ if and only if $\delta_\alpha\lConfig{l_{-1}}{l_0}{l_1} = \fState_\alpha$, and
   (3) $h_\mathtt{s}\lConfig{\qState_\alpha}{\qState_\alpha}{\qState_\alpha} = h_\mathtt{s}\lConfig{\qState_\alpha}{\qState_\alpha}{\oState} = \qState_\beta$.
\end{definition}

\begin{proposition}
   Let $\alpha$ be a FSSP solution CA, $\beta$ a FSSP-candidate CA and $h$ a local simulation from $\alpha$ to $\beta$.
   If $\beta$ is a FSSP solution, then $h$ is FSSP-compliant.
\end{proposition}


The following proposition is at the same time not difficult once noted, but extremely surprising and useful: the simple constraints above are also ``totally characterizing'' and the previous implication is in fact an equivalence.
This means in particular that it is not necessary to generate space-time diagrams to check if a constructed CA is an FSSP solution, which saves lot of computations.

\begin{proposition}
   Let $\alpha$ be an FSSP solution, $\beta$ a FSSP-candidate CA and $h$ be local simulation from $\alpha$ to $\beta$.
   If $h$ is FSSP-compliant, then $\beta$ is a FSSP solution.
\end{proposition}

\section{Exploring The Graph of Local Mappings}
\label{optim-problem}

\subsection{The Graph of FSSP-compliant Local Mappings}

In our actual algorithm, we take as input an existing FSSP solution $\alpha$ and fix a set of state $S$ of size $|\stateSet_\alpha|$.
The search space consists of all FSSP-compliant local mappings from $\alpha$ to $S$, the neighbors $\neighborhood(h)$ of a local mapping $h$ being all $h'$ that differs from $h$ on exactly one entry, \ie $\exists!(l_{-1},l_0,l_1) \in \dom(\transTab_\alpha) \text{ s.t. } h_\mathtt{s}(l_{-1},l_0,l_1) \neq h'_\mathtt{s}(l_{-1},l_0,l_1)$.
More precisely, the mappings are considered modulo bijections of $S$.
Indeed, two mappings $h$ and $h'$ are considered equivalent if there is some bijection $r: S \to S$ such that $h_\mathtt{z} = r \circ h'_\mathtt{z}$ and $h_\mathtt{s} = r \circ h'_\mathtt{s}$.
So the search space is, in a sense, made of equivalence classes, each class being represented by a particular element.
This element is chosen to be the only mapping $h$ in the class such that $h_\mathtt{s}$ is monotonic according to arbitrary total orders on $\dom(\transTab_\alpha)$ and $S$ fixed for the entire run of the algorithm.

Considering 6-states solutions, let us denote $\stateSet_\alpha = \{\oState_\alpha, \gState_\alpha, \qState_\alpha, \fState_\alpha, \aState_\alpha, \bState_\alpha, \cState_\alpha\}$ and $S = \{\oState, \gState, \qState, \fState, \aState, \bState, \cState\}$\footnote{Recall that we do not count the $\star$ states.}.
In each of these sets, four of the states are the special FSSP solution states (Def.~\ref{fssp_ready} and Def.~\ref{min_fssp_solution}).
Only the three states $\aState$, $\bState$, $\cState$ come with no constraints.
We can thus evaluate the size of the search space by looking at the degrees of freedom of FSSP-compliant local mappings (Def~\ref{fssp_compliant}).

Indeed, all FSSP-compliant local mappings $h$ from $\alpha$ to $S$ have the same partial function $h_\mathtt{z}$, and the same value $h_\mathtt{s}(l_{-1},l_0,l_1)$ for those entries $(l_{-1},l_0,l_1) \in \dom(\delta_\alpha)$ forced to $\star$, $\qState$ or $\fState$.
For all other entries $(l_{-1},l_0,l_1)$, $h_\mathtt{s}(l_{-1},l_0,l_1)$ cannot take the values $\star$ nor $\fState$, leaving 5 values available.
So given an initial solution $\alpha$, the number of local mappings is $5^x$ where $x$ is the size of $\dom(\transTab_\alpha)$ without those entries constrained in Def.~\ref{fssp_compliant}.
To give an idea, for the solution 668 of the 718 solutions, $x = 86$ to the size of the search has 61 digits, and for Mazoyer's solution, $x = 112$ leading to a number with 79 digits.


\subsection{Preparation Before the Algorithm}

As described in Section~\ref{sltt}, the local mappings are evaluated from the super local transition table.
To build this table, we generate, for each size $n$ from 2 to 5000, the space-time diagram $\fDiagram_\alpha(\bar{n})$ and collect all super local transitions occurring from time $0$ to $2n-4$ and from position $1$ to $n$.
Note that for all known minimum-time 6-state solutions, no additional super local transitions appear after $n = 250$.

The starting point of the exploration is the local mapping $h_\alpha$ corresponding to the local transition function $\transTab_\alpha$ itself, \ie $(h_\alpha)_\mathtt{z} = q \restriction \{ \oState_\alpha, \gState_\alpha, \qState_\alpha \}$ and $(h_\alpha)_\mathtt{s} = q \circ \transTab_\alpha$ for some bijection $q : \stateSet_\alpha \to S$.
This local mapping is obviously FSSP-compliant since it is local simulation from $\alpha$ to $\alpha$.

\subsection{The Exploration Algorithm}

\begin{figure}[h]
\SetInd{1em}{0.1em}
\noindent
\begin{minipage}[t]{.53\linewidth}
\begin{algorithm}[H]
    \Indp
    \Indm\explore{$\superTransTab_\alpha, h_\alpha, k$}\\
    \Indp
    $H \gets \{ h_\alpha \}$\\
    $H_{current} \gets \{ h_\alpha \}$\\
    \While{$\card{H_{current}}\ > 0$}{
        $H_{next} \gets \{ \}$\\
        \For{$h \in H_{current}$}{
            $S, H \gets \pertN(\superTransTab_\alpha, H, h, k)$\\
            \For{$h' \in (S \setminus H)$}{
                \If{$\isSimul(h', \superTransTab_\alpha)$}{
                    $H_{next} \gets H_{next} \cup \{h'\}$\\ 
                    $H \gets H \cup \{h'\}$
                }
            }
        }
        $H_{current} \gets H_{next}$
    }
    \Return $H$
    \caption{}
    \label{local-search-algo}
\end{algorithm}
\end{minipage}
\hfill
\begin{minipage}[t]{.47\linewidth}
\begin{algorithm}[H]
    \Indp
    \Indm $\pertN(\superTransTab_\alpha, H, h, k)$\\
    \Indp
            $S \gets \neighborhood(h)$\\
            $h' \gets \perturbation(h, k)$\\
            \If{$h' \not\in H$}{
                $S \gets S \union \neighborhood(h')$ \\
                \If{$\isSimul(h', \superTransTab_\alpha)$}{
                    $H \gets H \union \{h'\}$
                } 
            } 
    \Return $S$, $H$
    \caption{}
\end{algorithm}
\end{minipage}
\end{figure}

To explain the algorithm, let us first consider the last parameter $k$ to be $0$, so that line 7 of the $\explore$ algorithm can be considered to be simply $S \gets \neighborhood(h)$.
In this case, the algorithm starts with $h_\alpha$, and explores its neighbors to collect all local simulations, then the neighbors of those local simulations to collect more local simulations, and so on so forth until the whole connected components of the sub-graph consisting only of the local simulations in collected.

More precisely, the variable $H$ collects all local simulations, $H_{current}$ contains the simulation discovered in the previous round and whose neighbors should be examined in current round, and the newly discovered local simulations are put in $H_{next}$ for the next round.
The function $\isSimul$ uses the super local transition table to construct the local transition relation of $\Phi_h$ and check if it is functional, \ie if it is a local transition function of a valid CA ${\rm \Gamma}_{\Phi_h}$.
By our construction, a valid CA is necessarily an FSSP solution making this operation really cheap.

When $k > 0$, the neighborhood operation is altered to add more neighbors.
A local mapping obtained by $k$ modifications is considered and its neighborhood is added to the original the normal neighborhood, in the hope of discovering another connected component of the local simulation subgraph.

\section{Analyzing the Results}
\label{result}

\subsection{Analyzing the 718 solutions}

As mentioned in the introduction, this study began by the desire to analyse the 718 solutions found in~\cite{DBLP:journals/asc/ClergueVF18}.
These solutions, numbered from 0 to 717, are freely available online.
We tried to search local simulation relation between them as done in~\cite{DBLP:journals/jca/NguyenM20}.
Firstly, we found a slight mistake since there are 12 pairs of equivalent solutions up to renaming of states: 
(105, 676), (127, 659), (243, 599), (562, 626), (588, 619), (601, 609), (603, 689), (611, 651), (629, 714), (663, 684), (590, 596) and (679, 707).
This means that there are really 706 solutions, but we still refer to them as the 718 solutions with their original numbering.

Once local simulation relation established between the 718 solutions, we analyzed in the number of connected components and found 193 while expecting only a few.
When there is a local simulation $h$ from a CA $\alpha$ and a CA $\beta$, the number of differences between $h_\alpha$ and $h$ varies a lot, but the median value is 3.

\subsection{Analyzing the Local Simulations}

To find more FSSP solutions, we implemented many algorithms, gradually simplifying them into the one presented in this paper.
It has been run on an Ubuntu Marvin machine with 32 cores of 2.00GHz speed and 126Gb of memory.
However, the implementation being sequential, only two cores was used by the program.
The original plan was to generate as many solutions as possible but we had some problems with the management of quotas in the shared machine.
So we only expose the some selected data to show the relevance of the approach.

When running the program with the solution 355 and $k = 0$, the program used 14Gb of memory and stopped after 27.5 hours and found 9,584,134 local simulations!
A second run of the program for this solution with $k = 3$ found 11,506,263 local simulations after 80.5 hours.
This indicates that perturbations are useful but the second run find only 1922129 additional local simulations but its computing time is three times more than the first run.
Testing whether a local simulation belongs to set $H$ obviously takes more and more times as more local mappings are discovered but there might be some understanding to gain about the proper mapping landscape too in order to improve the situation.

The transition table for the original Mazoyer's solution can be found in~\cite{DBLP:journals/tcs/Mazoyer87}, but also in~\cite{DBLP:journals/ijuc/UmeoHS05} together with other minimal-time solution transition table.
When running the program of the original Mazoyer's solution with different values of $k$ with obtained the following number of new solutions for different runs.
The behavior with $k=1$ seems to be pretty robust, but the bigger number of results is obtained with $k = 2$. 

\begin{center}
\begin{tabular}{|l|l|}
    \hline
    k & number of solutions found by 10 different runs \\
    \hline \hline
    0 & \textbf{644} \\
    \hline
    1 & 20682, 17645, 20731, 16139, \textbf{20731}, 9538, 20626, 20682, 20054, 20490 \\
    \hline
    2 & 9451, 9451, 20595, 8241, \textbf{37275}, 3817, 17421, 8241, 17317, 19895 \\
    \hline
    3 & 644, 644, 644, 644, 644, 644, 644, 731, 8241, \textbf{8241} \\
    \hline
    4 & 644, 644, 644, 2908, 644, 644, 644, 644, 644, \textbf{8241} \\
    \hline
    5 & 644, 644, 644, 644, 644, 644, 644, 644, 644, 644 \\
    \hline
\end{tabular}
\end{center}

\begin{figure}
\centering
\begin{subfigure}{.45\textwidth}
\centering
\includegraphics[width=\textwidth]{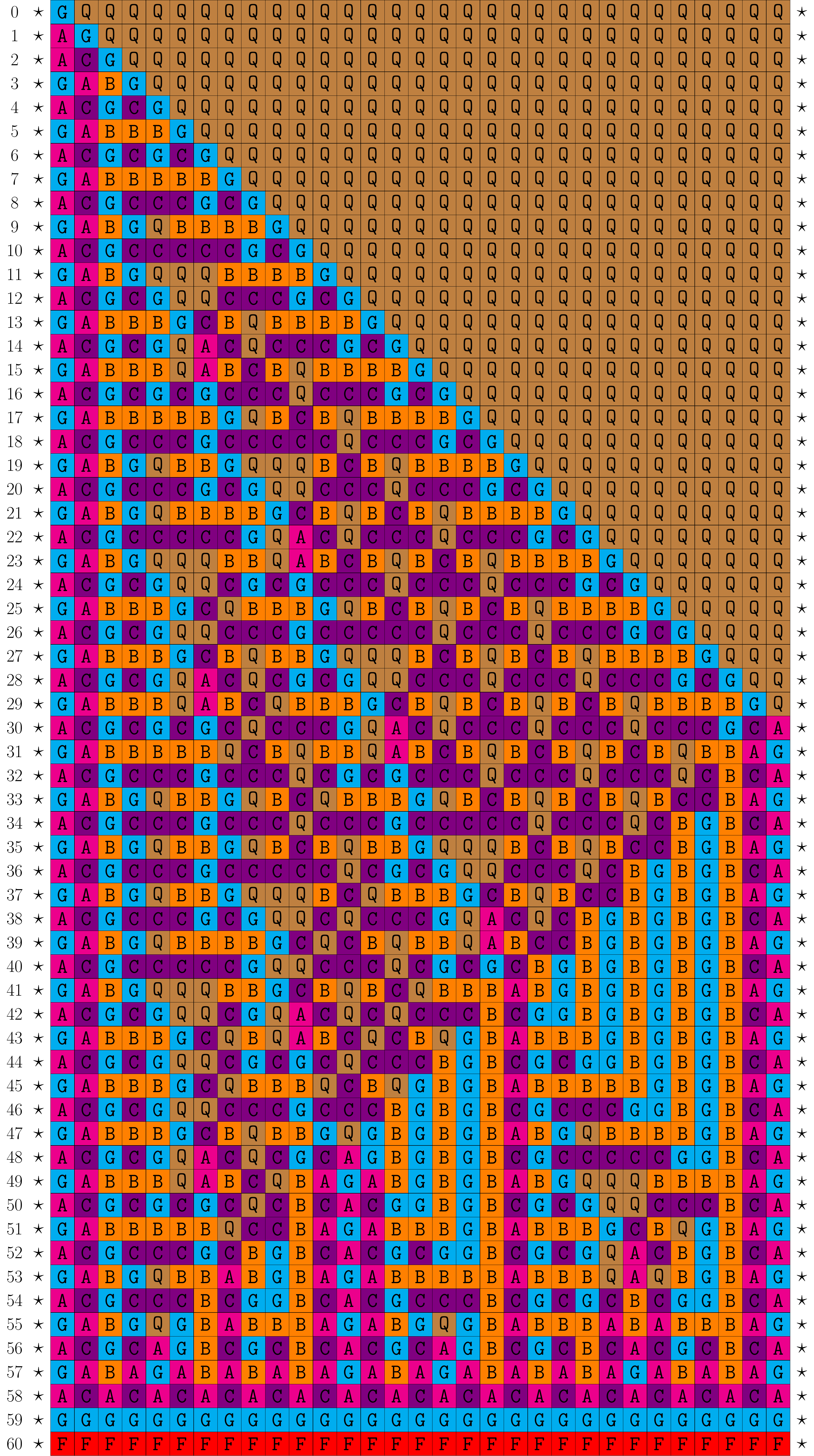}
\caption{original solution 668: 93 rules}
\label{fig:668_origine}\label{fig:two-third}
\end{subfigure}
\hspace*{1cm}
\begin{subfigure}{.45\textwidth}
\centering
\includegraphics[width=\textwidth]{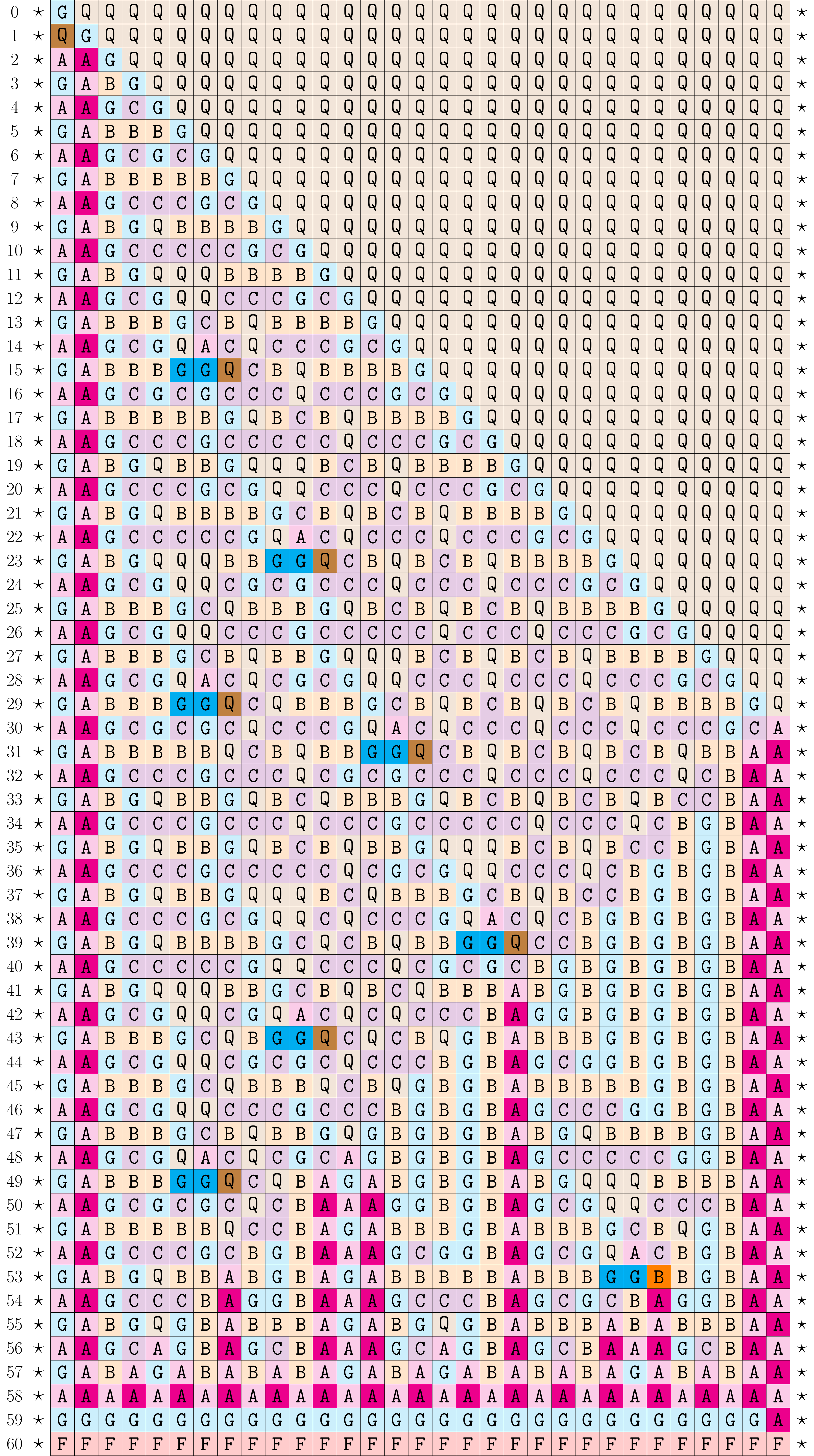}
\caption{a local simulation of 668: 90 rules}
\label{fig:668_locSim}
\end{subfigure}
\vspace{.1cm}
\newline

\begin{subfigure}{.45\textwidth}
\centering
\includegraphics[width=\textwidth]{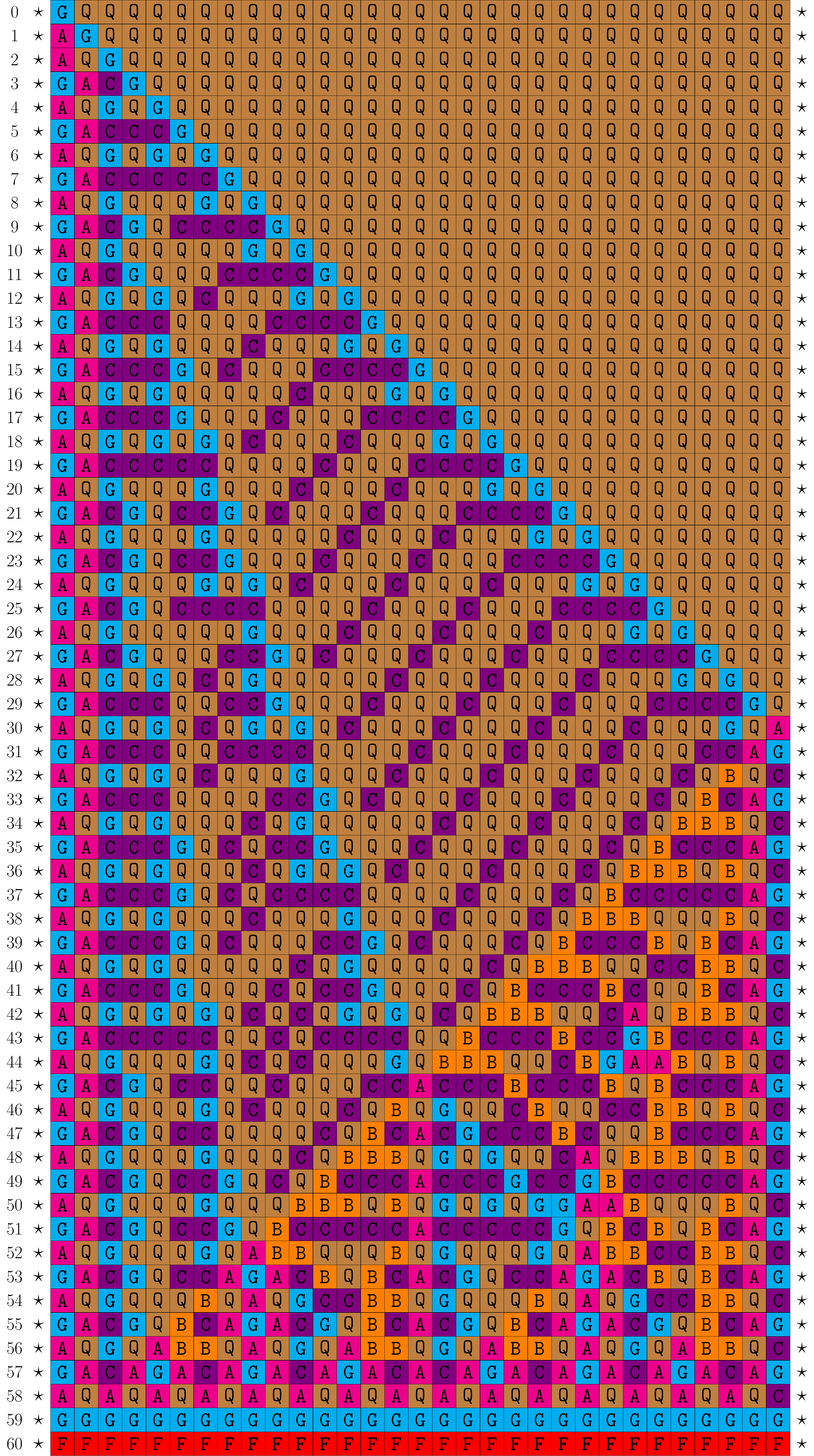}
\caption{original solution 355}
\label{fig:355_origine}\label{fig:half}
\end{subfigure}
\hspace*{1cm}
\begin{subfigure}{.45\textwidth}
\centering
\includegraphics[width=\textwidth]{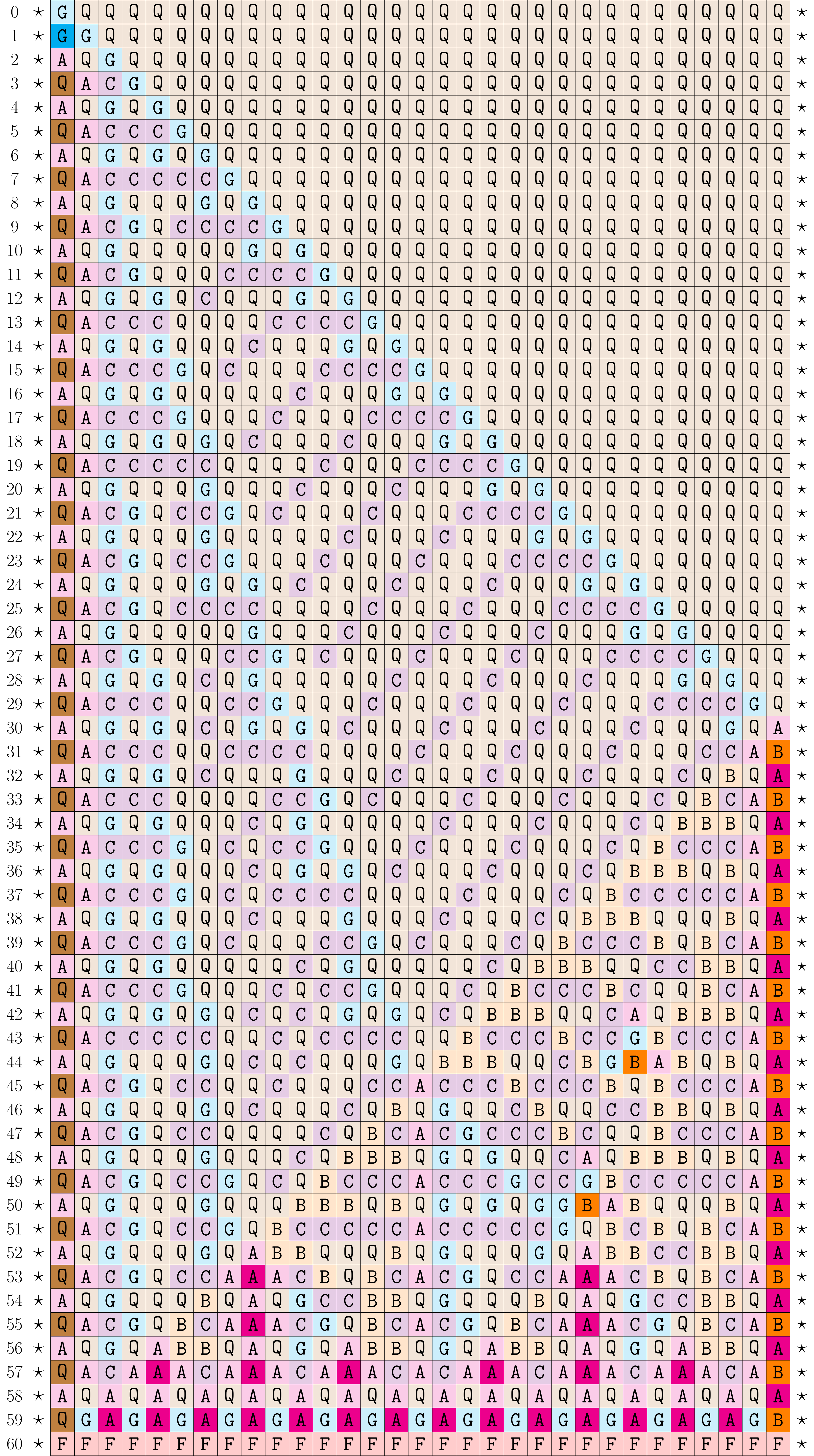}
\caption{a local simulation of 355}
\label{fig:355_locSim}
\end{subfigure}
\caption{Some FSSP space-time diagrams of size 31}
\label{fig:dgm}
\end{figure}

Note that while the solutions do not have less states, the number of transitions do change.
We show in Figure~\ref{fig:668_origine} the solution 668 (the only Mazoyer-like solutions found among the 718 solutions), and one of its simulations having less transition in Figure~\ref{fig:355_origine}.
For fun, we also show in Figure~\ref{fig:355_locSim} and~\ref{fig:668_locSim} a local simulation having alternating states at time $2n-3$, illustrating how local simulation rearrange locally the information.
The identical part in represented with lighter colors to highlight the differences.

\begin{proposition}
   There are at least many millions of minimum-time 6-state FSSP solutions.
\end{proposition}

\section{Conclusion}
\label{conclusion}

This paper presents only a small part of many ongoing experimentations.
The notion of local simulation presented here is just a particular case of the notion of cellular field that can be used more broadly to investigate these questions.
For example, we relate here only the small cones $\{d(t-dt,p+dp) \mid dt \in \{0,1\}, dp \in \itv{-dt}{+dt} \}$ of any space-time diagram $d$ in local mappings.
If we increase the range of $dt$ in this definition to be $\itv{0}{h}$ for some $h > 1$, we allow CA to be transformed to a bigger extent.

Another justification for this extension is that the composition of two local simulations is not a local simulation.
In fact, composing an $h$-local simulation with an $h'$-local simulation produces an $(h+h')$-local simulation in general.
A 0-local simulation is just a (possibly non-injective) renaming of the states.

Note that since local mappings of local mappings are not local mappings, running the above algorithm on new found solutions should a priori generate more solutions!
Of course, a more exhaustive study is required.

Our guess is that, with a properly large notion of such simulations, it should be possible to classify the 718 solutions into only a few equivalence classes, more or less in two groups: the ``mid-way division'' solutions and the ''two-third division'' solutions.
This results also represents an important step in the understanding of automatic optimization of CA.


Finally, the content of Section~\ref{sec:correctness-preserving} about the preservation of correctness by FSSP-compliant local simulation is really interesting because of the simplicity of checking FSSP-compliance.
It implies that a proof of correctness of a small FSSP solution can indeed be made on some huge, possibly infinite, simulating CA where everything is explicit as considered in~\cite{DBLP:conf/acri/MaignanY14,DBLP:journals/jca/MaignanY16}.
This can be applied to ease the formal proof of correctness of Mazoyer's solution.
Up to our knowledge, it is known to be long and hard but also to be the only proof to be precise enough to actually be implemented in the Coq Proof Assistant~\cite{duprat:hal-02101837}.

We would like to give special thanks to Jean-Baptiste Yun\`es who pointed us the 718 solutions paper.
If we are right, he also partly inspired the work who lead to the 718 solutions by a discussion during a conference.

\bibliographystyle{plain}
\bibliography{reference}

\begin{thebibliography}{10}

\bibitem{DBLP:journals/iandc/Balzer67}
Robert Balzer.
\newblock An 8-state minimal time solution to the firing squad synchronization
  problem.
\newblock {\em Information and Control}, 10(1):22--42, 1967.

\bibitem{DBLP:journals/asc/ClergueVF18}
Manuel Clergue, S{\'{e}}bastien V{\'{e}}rel, and Enrico Formenti.
\newblock An iterated local search to find many solutions of the 6-states
  firing squad synchronization problem.
\newblock {\em Appl. Soft Comput.}, 66:449--461, 2018.

\bibitem{duprat:hal-02101837}
Jean Duprat.
\newblock {Proof of correctness of the Mazoyer's solution of the firing squad
  problem in Coq}.
\newblock Research Report LIP RR-2002-14, {Laboratoire de l'informatique du
  parall{\'e}lisme}, March 2002.

\bibitem{10012108193}
H.~D. Gerken.
\newblock Uber synchronisations-probleme bei zellularautomaten.
\newblock {\em Diplomarbeit, Institut fur Theoretische Informatik, Technische
  Universitat Braunschweig}, 50, 1987.

\bibitem{DBLP:conf/acri/MaignanY12}
Luidnel Maignan and Jean{-}Baptiste Yun{\`{e}}s.
\newblock A spatio-temporal algorithmic point of view on firing squad
  synchronisation problem.
\newblock In Georgios~Ch. Sirakoulis and Stefania Bandini, editors, {\em
  Cellular Automata - 10th International Conference on Cellular Automata for
  Research and Industry, {ACRI} 2012, Santorini Island, Greece, September
  24-27, 2012. Proceedings}, volume 7495 of {\em Lecture Notes in Computer
  Science}, pages 101--110. Springer, 2012.

\bibitem{DBLP:conf/acri/MaignanY14}
Luidnel Maignan and Jean{-}Baptiste Yun{\`{e}}s.
\newblock Experimental finitization of infinite field-based generalized {FSSP}
  solution.
\newblock In Jaroslaw Was, Georgios~Ch. Sirakoulis, and Stefania Bandini,
  editors, {\em Cellular Automata - 11th International Conference on Cellular
  Automata for Research and Industry, {ACRI} 2014, Krakow, Poland, September
  22-25, 2014. Proceedings}, volume 8751 of {\em Lecture Notes in Computer
  Science}, pages 136--145. Springer, 2014.

\bibitem{DBLP:journals/jca/MaignanY16}
Luidnel Maignan and Jean{-}Baptiste Yun{\`{e}}s.
\newblock Finitization of infinite field-based multi-general {FSSP} solution.
\newblock {\em J. Cellular Automata}, 12(1-2):121--139, 2016.

\bibitem{DBLP:journals/tcs/Mazoyer87}
Jacques Mazoyer.
\newblock A six-state minimal time solution to the firing squad synchronization
  problem.
\newblock {\em Theor. Comput. Sci.}, 50:183--238, 1987.

\bibitem{DBLP:journals/jca/NguyenM20}
Tien~Thao Nguyen and Luidnel Maignan.
\newblock Some cellular fields interrelations and optimizations in {FSSP}
  solutions.
\newblock {\em J. Cellular Automata}, 15(1-2):131--146, 2020.

\bibitem{DBLP:journals/tcs/Noguchi04}
Kenichiro Noguchi.
\newblock Simple 8-state minimal time solution to the firing squad
  synchronization problem.
\newblock {\em Theor. Comput. Sci.}, 314(3):303--334, 2004.

\bibitem{DBLP:conf/parcella/Sanders94}
Peter Sanders.
\newblock Massively parallel search for transition-tables of polyautomata.
\newblock In Chris~R. Jesshope, Vesselin Jossifov, and Wolfgang Wilhelmi,
  editors, {\em Parcella 1994, {VI.} International Workshop on Parallel
  Processing by Cellular Automata and Arrays, Potsdam, Germany, September
  21-23, 1994. Proceedings}, volume~81 of {\em Mathematical Research}, pages
  99--108. Akademie Verlag, Berlin, 1994.

\bibitem{DBLP:journals/amc/UmeoHNIS18}
Hiroshi Umeo, Mitsuki Hirota, Youhei Nozaki, Keisuke Imai, and Takashi Sogabe.
\newblock A new reconstruction and the first implementation of goto's {FSSP}
  algorithm.
\newblock {\em Appl. Math. Comput.}, 318:92--108, 2018.

\bibitem{DBLP:journals/ijuc/UmeoHS05}
Hiroshi Umeo, Masaya Hisaoka, and Takashi Sogabe.
\newblock A survey on optimum-time firing squad synchronization algorithms for
  one-dimensional cellular automata.
\newblock {\em {IJUC}}, 1(4):403--426, 2005.

\bibitem{DBLP:journals/iandc/Waksman66}
Abraham Waksman.
\newblock An optimum solution to the firing squad synchronization problem.
\newblock {\em Information and Control}, 9(1):66--78, 1966.

\end{thebibliography}

\end{document}